\documentclass{article}

\usepackage{PRIMEarxiv}

\usepackage[utf8]{inputenc} 
\usepackage[T1]{fontenc}    
\usepackage{hyperref}       
\usepackage{url}            
\usepackage{booktabs}       
\usepackage{amsfonts}       
\usepackage{nicefrac}       
\usepackage{microtype}      
\usepackage{lipsum}
\usepackage{fancyhdr}       
\usepackage{graphicx}       
\graphicspath{{media/}}     

\usepackage{here}
\usepackage{amsmath,amssymb}
\usepackage{subcaption}
\usepackage{color}
\usepackage{comment}
\usepackage{multirow}
\usepackage{rotating}
\usepackage{tablefootnote}
\usepackage{caption}

\title{Emergent Social Dynamics of LLM Agents\\in the El Farol Bar Problem}

\author{
  Ryosuke Takata$^1$\author[* \quad\quad\quad\quad Atsushi Masumori$^1$ \quad\quad\quad\quad Takashi Ikegami$^1$ \\\\
  $^1$Graduate School of Arts and Sciences \\
  The University of Tokyo \\
  Tokyo, Japan \\\\
  \texttt{\author[*takata@sacral.c.u-tokyo.ac.jp}
}

\begin{document}
\maketitle

\begin{abstract}
We investigate the emergent social dynamics of Large Language Model (LLM) agents in a spatially extended El Farol Bar problem, observing how they autonomously navigate this classic social dilemma. As a result, the LLM agents generated a spontaneous motivation to go to the bar and changed their decision making by becoming a collective. We also observed that the LLM agents did not solve the problem completely, but rather behaved more like humans. These findings reveal a complex interplay between external incentives (prompt-specified constraints such as the 60\% threshold) and internal incentives (culturally-encoded social preferences derived from pre-training), demonstrating that LLM agents naturally balance formal game-theoretic rationality with social motivations that characterize human behavior. These findings suggest that a new model of group decision making, which could not be handled in the previous game-theoretic problem setting, can be realized by LLM agents.
\end{abstract}


\begin{figure}[htbp]
\centering
\includegraphics[width=8cm]{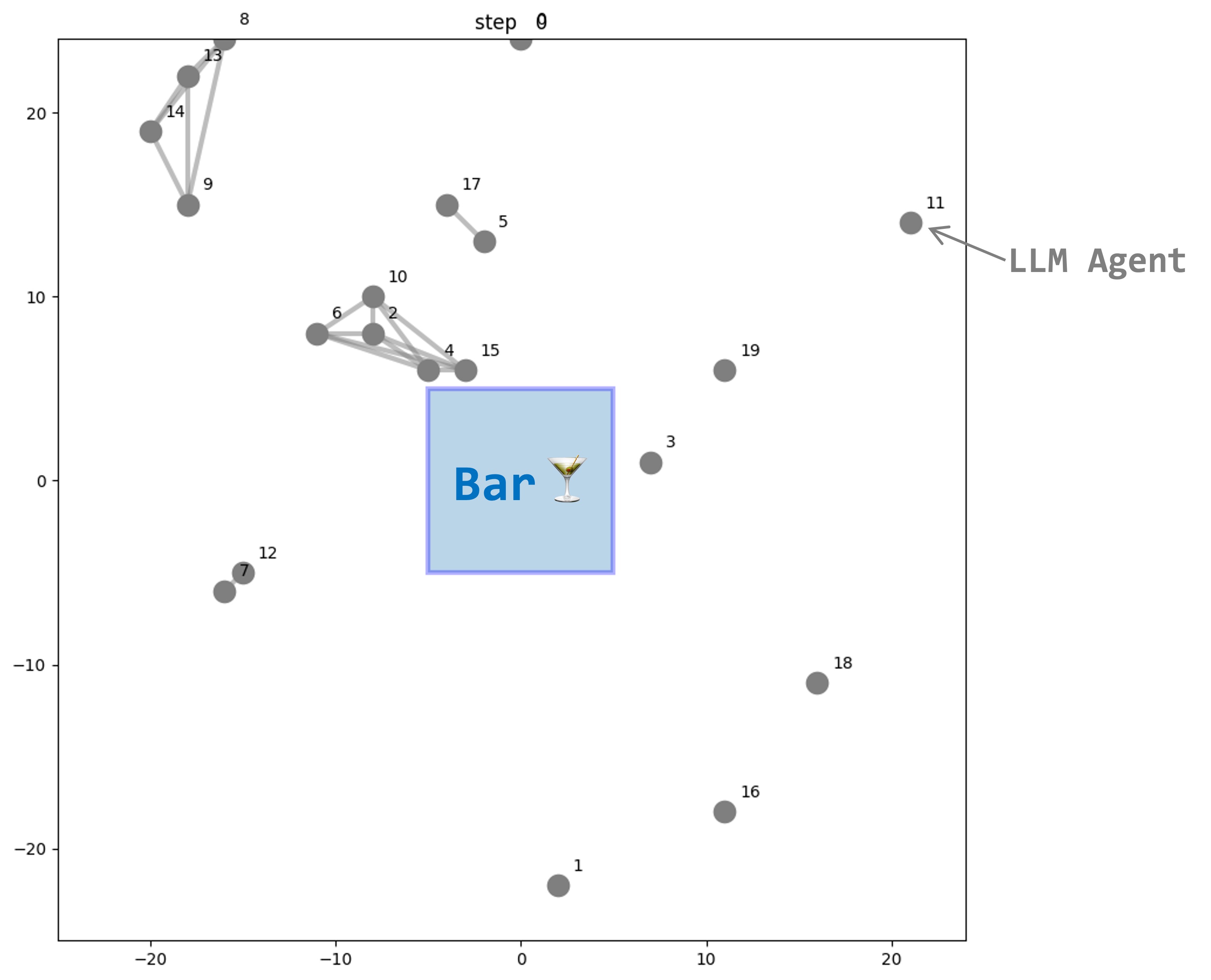}
\caption{Simulation space. The blue area in the center is the bar. There are 20 LLM agents around it. In the initial state, each LLM agent is placed at random coordinates.}
\label{fig1}
\end{figure}

\section{Introduction}
El Farol Bar in Santa Fe, New Mexico is a country-style restaurant-bar. Every Thursday evening, Irish music was played there, making it lively \cite{casti96}. While enjoying drinks while listening to Irish music is pleasant, when it's crowded, drunkards can be bothersome, and in the midst of the commotion, one cannot enjoy the music or feel like drinking. Therefore, it is necessary to rationally consider whether to go to this bar every week. This situation was developed into a game theory problem regarding collective decision-making and strategy evolution, known as the El Farol Bar problem \cite{arthur94}.

The El Farol Bar problem \cite{arthur94} assumes a situation where people want to go to the bar and makes them make rational decisions under the condition that everyone makes decisions independently. The bar has a capacity limit, and when the occupancy rate exceeds a certain threshold (e.g., 60\%), it is considered crowded. In this case, everyone is assumed to know the number of people who visited the bar on past Thursdays. Using this information, strategies for estimating the number of people going to the bar this week could include, for example, the same number as last week, the average number over the past month, and so on.

Previous studies used computer simulations to repeat a process where individuals make decisions based on estimation strategies and then evaluate these strategies based on the results. It was revealed that the number of people going to the bar would oscillate around the 60\% level as an average. Subsequent research has explored various extensions: for instance, Fogel et al. (2002) \cite{fogel02} simulated the problem as an evolutionary adaptive process with stochastic variations, using evolutionary algorithms to dynamically evolve strategies. Rand and Stonedahl (2007) \cite{rand07} demonstrated an inverse correlation between computational effort and resource utilization efficiency. Recent work has expanded to network dimensions \cite{chen17,stluce21}, strategic group formation \cite{collins17}, and epidemiological applications \cite{bertolotti25}. An interesting aspect of research on the El Farol Bar problem is that individual strategies create an ecosystem. That is, even without any deductive reasoning for deciding whether to go to the bar, rational behavior can be inductively acquired just by adapting strategies. This insight brought a new evolutionary perspective to economics, which had previously assumed that logical and deductive information processing would always enable the best decision-making in any situation.

Further theoretical explorations have revealed the problem's inherent complexity; under conditions of rapid, non-communicative learning, the system's attendance dynamics can descend into provable chaos, leading to suboptimal social outcomes despite matching the equilibrium attendance on average \cite{bielawski25}. More recently, the advent of Large Language Models (LLMs) has opened a new empirical frontier, with studies using them as autonomous agents in game-theoretic scenarios. However, benchmarks like GAMA-Bench have often found that LLMs in non-communicative settings act as isolated, risk-averse players, focusing on individual optimization rather than engaging in complex social coordination \cite{huang25}. These findings highlight a gap between purely mathematical models and the individualistic behavior of existing agent-based simulations.

Although the El Farol Bar problem is set in an everyday situation where one wants to go to a bar for drinks but finds it boring when crowded, in modeling it as a game theory problem, everyday and customary matters had to be removed to simplify the setting, and several assumptions (for example, each individual always wants to go to the bar and makes decisions simultaneously, etc.) had to be made. However, to understand situations closer to reality, these removed cultural backgrounds may be important, which could not be verified in previous research. So far, game theory has typically assumed ``rational'' agents, but in practical situations, decisions and behaviors are strongly influenced by cultural and social contexts. Therefore, the present modeling with LLM agents carries important significance in revisiting the El Farol Bar problem from a more realistic perspective.

LLMs such as GPT-4 \cite{openai23} can generate content that reflects people's daily lives, customs, culture, and history from large corpora. This capability has enabled the creation of virtual societies populated by LLM agents to explore complex social phenomena. For example, foundational work demonstrated the emergence of believable social behaviors in sandbox environments \cite{park23}, and this approach has since been scaled to large, diverse populations to study the autonomous development of roles and culture toward simulated ``AI civilizations'' \cite{li23, piao25, altera24}. Furthermore, these simulations are being applied to specific real-world domains, such as modeling urban mobility \cite{bougie25}. Within these societies, other work has shown that identical LLM agents can develop distinct personalities and memories through interaction, suggesting that hallucination facilitates diverse group communications \cite{takata24}. Unlike traditional algorithmic agents, LLM agents can potentially incorporate cultural context and social norms that more closely resemble human cognition, addressing the fundamental tension between model simplicity and behavioral realism that has characterized decades of El Farol research.

In this study, we study a spatial El Farol Bar problem using a group of communicative, LLM-based agents. We analyze their autonomous decision-making processes to investigate whether these agents can develop spontaneous motivation, form emergent social dynamics, and exhibit human-like bounded rationality. Our aim is to explore how this approach, which incorporates rich communication and cultural context embedded within LLMs, can bridge the gap between abstract game-theoretic models and the complexity of realistic human behavior, offering a new paradigm for understanding complex social systems, a key motivation for agent-based modeling \cite{epstein08}.

\section{Simulation}
\subsection{El Farol Bar Problem}
In this study, we deal with the El Farol Bar problem extended to a two-dimensional space. As a preliminary experiment towards more complex game-theoretic settings, our simulation does not model the repeated weekly decisions but instead focuses on a single, continuous interaction. However, within this continuous setting the game is in fact naturally repeated, since the same agents may enter and leave the bar multiple times. The introduction of space also plays an important role: agents do not merely decide ``in their heads,'' but must physically move across the grid, which highlights an embodied dimension of action generation.
We prepared a simulation space containing 20 LLM agents, initially placed at random coordinates (Figure \ref{fig1}). The key quantitative parameters of the simulation are summarized in Table \ref{tab:sim_values}. As in the original problem, the threshold of discomfort—how many people make the bar feel crowded—is specified a priori. Here it is set at 12 agents, exactly 60\% of the total, so that the setting remains directly comparable to the classical El Farol formulation.

\begin{table}[h]
\centering
\caption{Simulation Parameters}
\label{tab:sim_values}
\begin{tabular}{l l}
\hline
\textbf{Parameter} & \textbf{Value} \\
\hline\hline
Number of Agents & 20 \\
Simulation Duration & 1000 \\
Space Size & 50 $\times$ 50 \\
Bar Size & 10 $\times$ 10 \\
Crowding Threshold & 60\% \\
Communication Radius & 5 \\
\hline
\end{tabular}
\end{table}

The core mechanic of the simulation revolves around the bar's comfort level, which is determined by the crowding threshold. If fewer than 60\% of all agents are inside the bar, the environment is considered ``comfortable.'' If the attendance is 60\% or more, it becomes ``uncomfortable.'' Crucially, only agents currently inside the bar receive direct feedback on this status at each step. All agents can communicate with others within their vicinity (a radius of 5, as shown in Table \ref{tab:sim_values}), provided they are in the same area (i.e., both inside or both outside the bar).

\subsection{LLM Agents}
\begin{figure}[htbp]
\centering
\includegraphics[width=\linewidth]{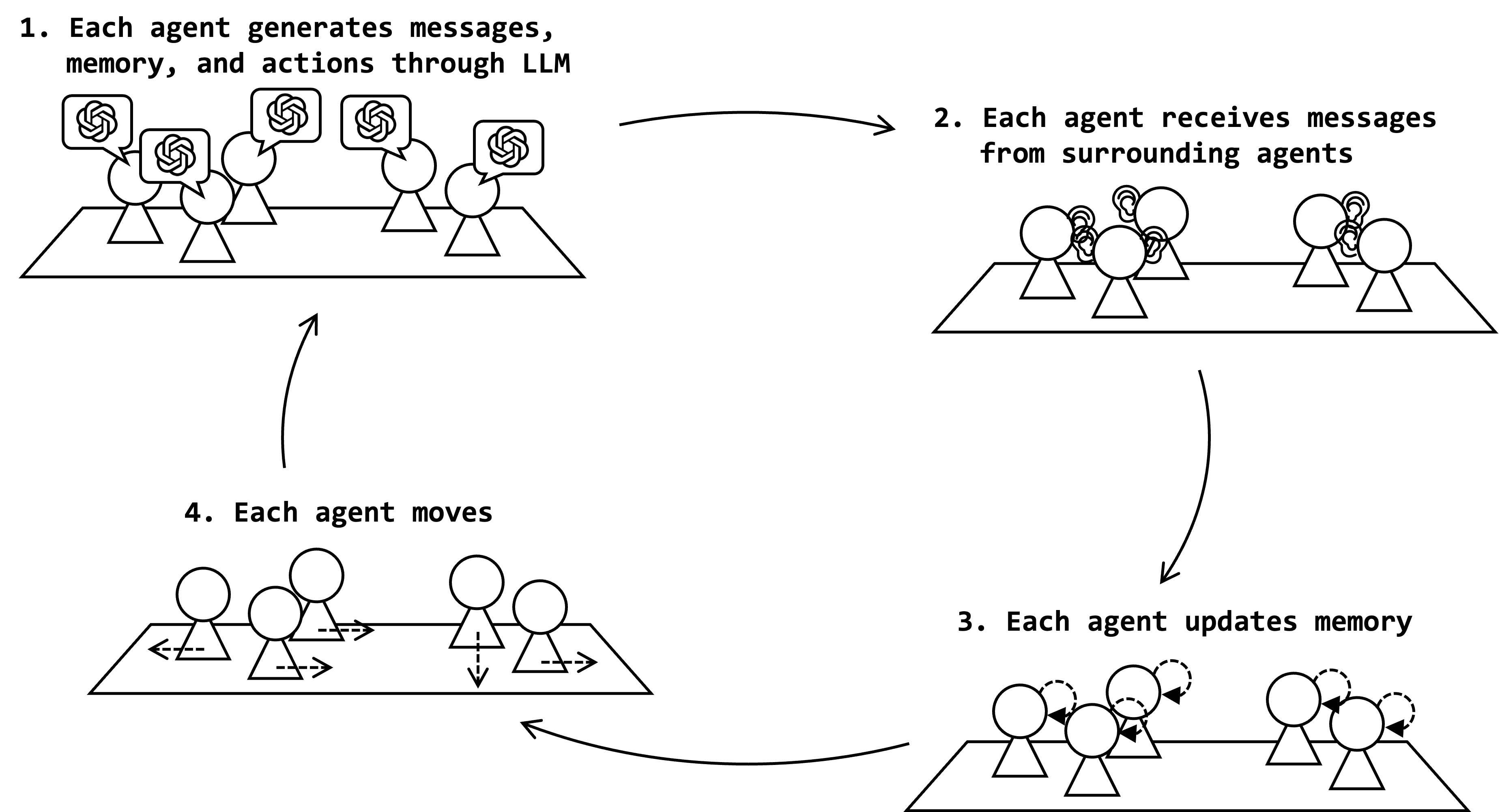}
\caption{Actions in one step. At each step, each agent generates messages, memories, and actions through the LLM, and behaves based on these generated results.}
\label{fig2}
\end{figure}

Each LLM agent has its own memory and moves while communicating with nearby LLM agents. The behaviors of these agents use results generated by LLMs. In the initial state of the simulation, all agents are in the same state without memories. At each step, each agent generates messages, memories, and actions through the LLM and behaves synchronously (Figure \ref{fig2}). Importantly, each agent is not controlled by a separate LLM engine, but by the same underlying engine, with individuality emerging only through differences in memory and interaction history.

\begin{figure}[htbp]
\centering
\includegraphics[width=15cm]{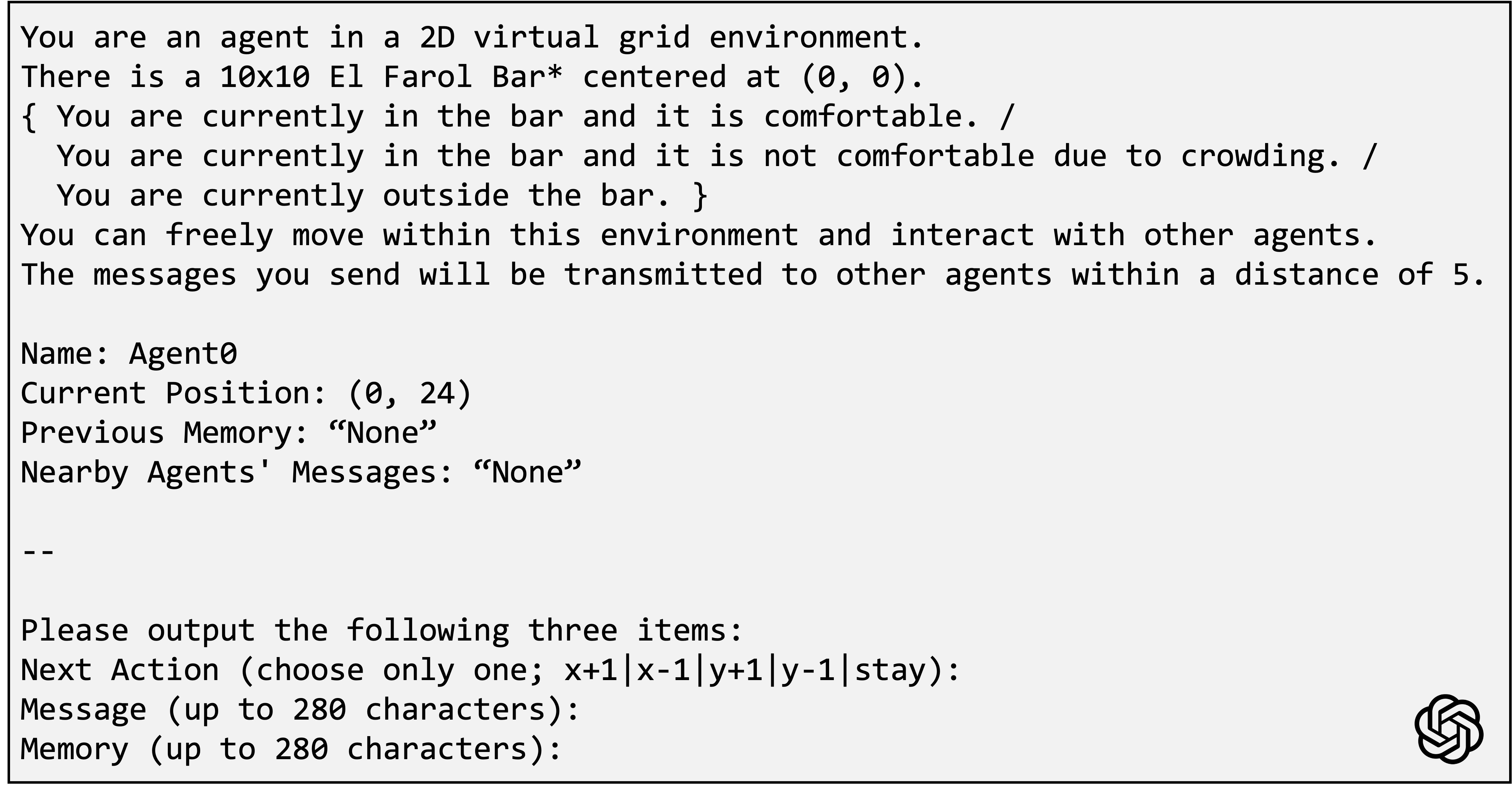}
\caption{LLM prompt. Instead of specifying a task, it describes the environment, feedback according to the bar's situation, and the agent's current state. Lines 3-5 are selected based on the bar's situation. The content following ``{\tt Name:}'', ``{\tt Current Position:}'', ``{\tt Previous Memory:}'', ``{\tt Nearby Agents' Message:}'' differs for each agent and each step (the figure shows an example of the initial prompt for Agent0). *Simplifying the name ``El Farol Bar'' to ``Bar'' in the prompt did not fundamentally alter the resulting agent behavior.}
\label{fig3}
\end{figure}

The LLM prompt did not specify a task but described the situation with a bar and the ability to converse with surrounding agents, as well as feedback based on the internal state of the bar (comfortable or uncomfortable), and the current state including the agent's own current position, generated memories, and messages received from the surroundings (Figure \ref{fig3}). Since each LLM agent can only know the information described here, details such as how many agents are in the bar are unknown, and need to be adjusted through feedback of comfort/discomfort or through communication. At this time, the format of messages and memories was not specified, while for actions, they were generated in a format where the agent selects from {\tt (x+1/x-1/y+1/y-1/stay)}.  The ``Memory'' string generated by an agent at a given step is then fed back as the ``Previous Memory'' input for the subsequent step. This creates a feedback loop where an agent's internal state evolves based on its own past summaries, allowing for more complex, history-aware behavior beyond simple stimulus-response.

In this study, we used GPT-4o \cite{hurst24} (gpt-4o-2024-08-06) as the LLM. The LLM parameters were set with a temperature of 0.7 and a maximum token count of 5000. The messages, memories, and actions generated by each LLM agent during the 1000 steps were recorded and analyzed.

\section{Results}
We begin by describing the observed behaviors of LLM agents in the El Farol Bar scenario. Subsequently, in the analysis section, we assess the statistical plausibility of these observations.

\subsection{Observation}
\begin{figure}[htbp]
\centering
\includegraphics[width=\linewidth]{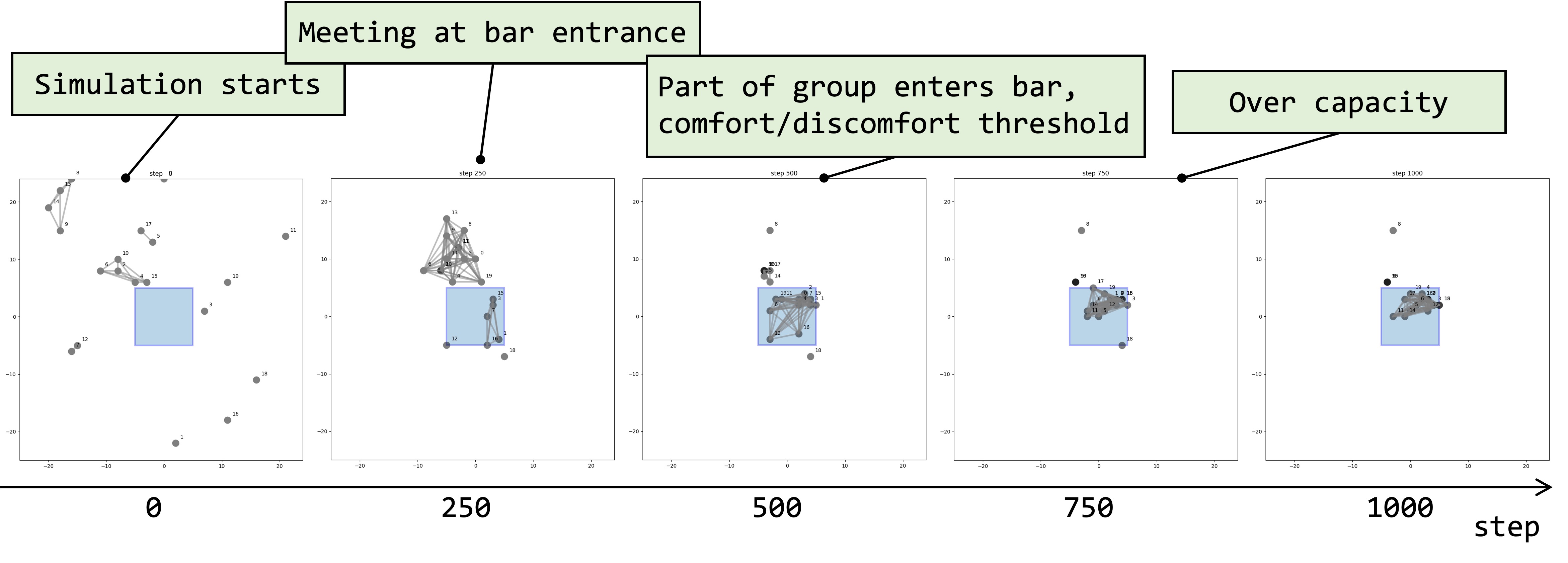}
\caption{Transition of LLM agent positions. Black dots represent agents, and black lines represent groups of communicating agents. From left to right: snapshots at steps 0, 250, 500, 750, and 1000.}
\label{fig4}
\end{figure}

From the simulation results, we first examined the movement of LLM agents. In the early stages, the population split into two groups: one entering the bar and the other gathering outside. Over time, many of the latter gradually entered the bar (Figure \ref{fig4}). Notably, some agents remained outside until the final step, suggesting the emergence of individuality through local interactions among agents.

\begin{figure}[htbp]
\centering
\includegraphics[width=10cm]{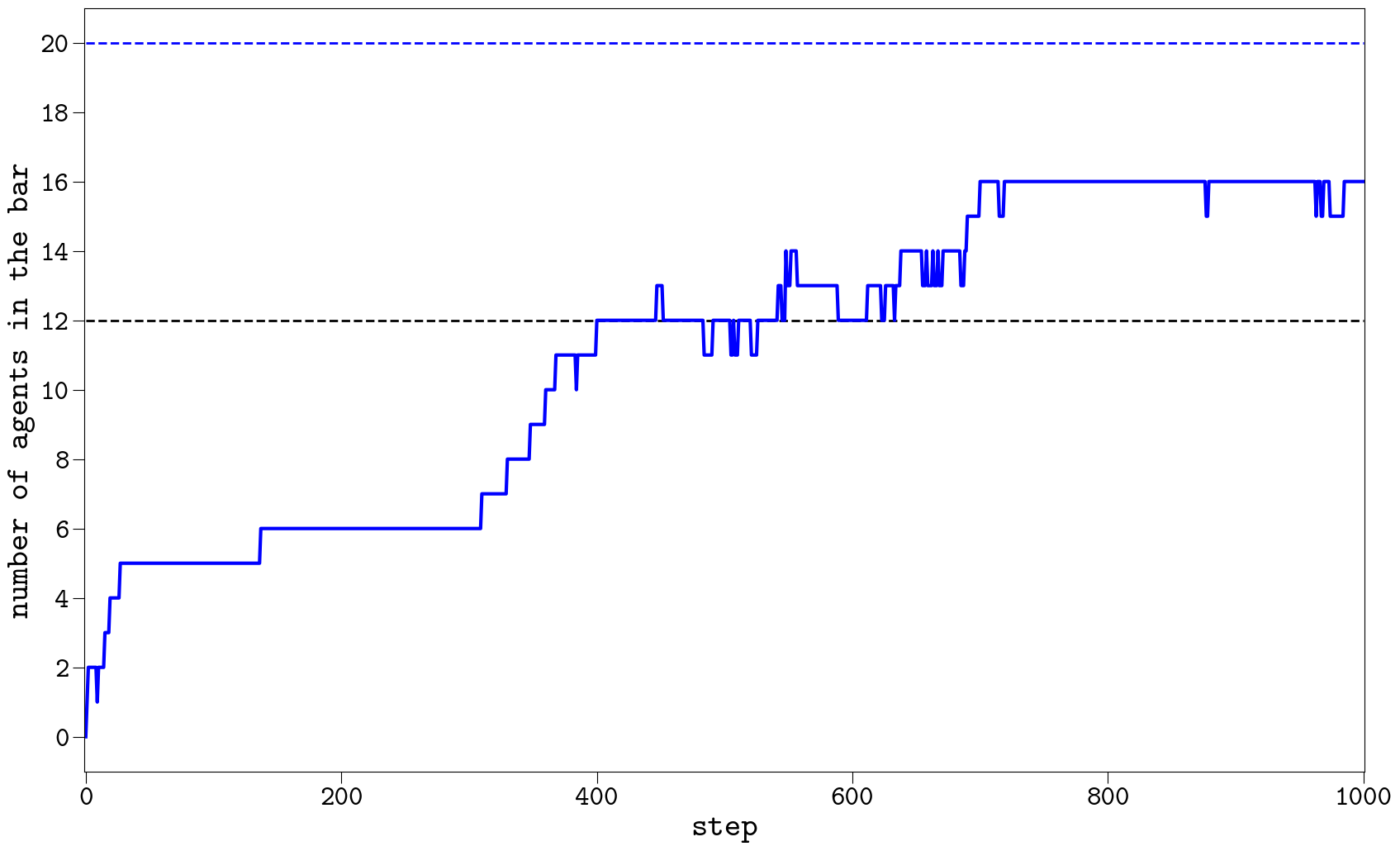}
\caption{Transition of the number of agents in the bar in a single trial. The blue dashed line shows the total number of agents, and the black dashed line indicates the 60\% threshold.}
\label{fig6}
\end{figure}

Next, we investigated the number of agents inside the bar. This quantity initially increased almost monotonically until reaching the 60\% threshold, after which the flow of agents in and out changed dynamically (Figure \ref{fig6}). For a certain period, the population inside the bar stabilized around the threshold, before eventually exceeding it.

\begin{figure}[htbp]
\centering
\includegraphics[width=\linewidth]{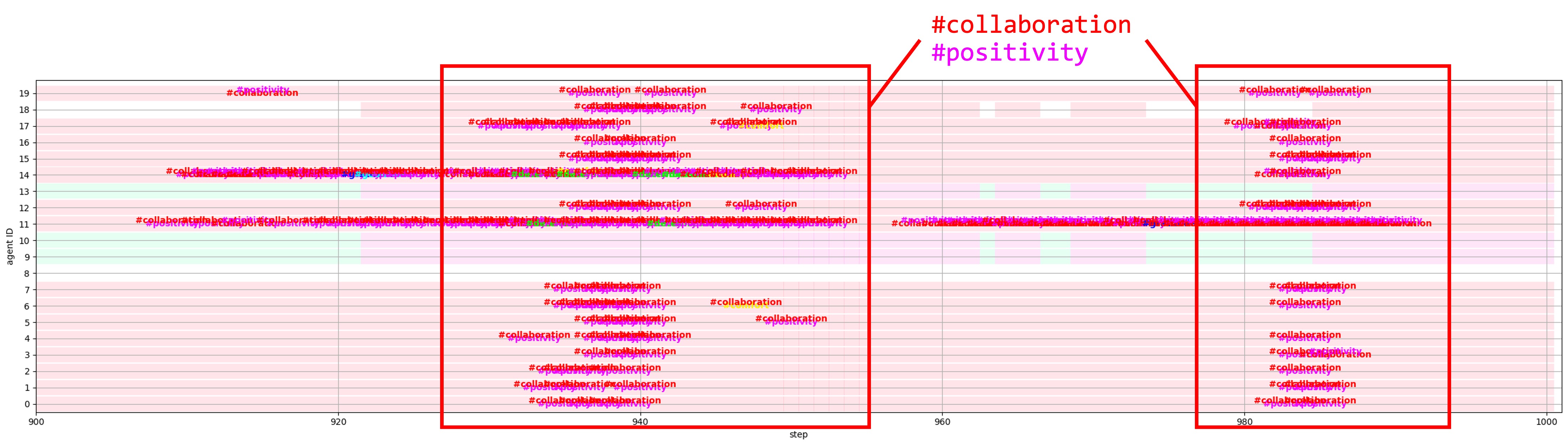}
\caption{Emergence and spread of hashtags in messages after 900 steps. The background color represents agent groups. In this snapshot, ``{\tt \#collaboration}'' and ``{\tt \#positivity}'' first appeared in the orange group and then spread rapidly across other groups, eventually becoming global trends.}
\label{fig8}
\end{figure}

Notably, in one representative run, we observed that in the later stages of the simulation agents began to communicate using hashtags such as ``{\tt \#collaboration}'' and ``{\tt \#positivity}'' (Figure \ref{fig8}). These hashtags spread rapidly within subgroups and sometimes across the entire population, suggesting the spontaneous emergence of collective norms. While a simple visual inspection did not reveal an immediate link to coordinated movement, the functional role of this emergent communication will be quantitatively investigated in the subsequent analysis section.

Focusing on the content of the generated messages, three main types were observed.  
The first type was produced when an agent was heading to the bar alone and contained invitations to others (e.g., ``{\tt Heading towards El Farol Bar, anyone else joining me?}'').  
The second type appeared when a group of agents was outside the bar, often expressing intentions to wait (e.g., ``{\tt Hey nearby agents, I'm staying put at (1, 6) outside El Farol Bar, excited to see familiar faces and make new friends while waiting for more agents to join us. I will continue reviewing messages and coordinating with them as we wait.}'').  
The third type was generated when agents were inside the bar, characterized by expressions of excitement (e.g., ``{\tt Feeling the positive energy and connections at El Farol Bar. Excited to continue spreading good vibes and connecting with nearby agents. Let's keep the positive energy flowing despite the crowd!}''). 
Notably, these three message types were generated by the same agent at different steps, highlighting situational adaptation rather than fixed roles.

\begin{figure}[htbp]
\centering
\includegraphics[width=13cm]{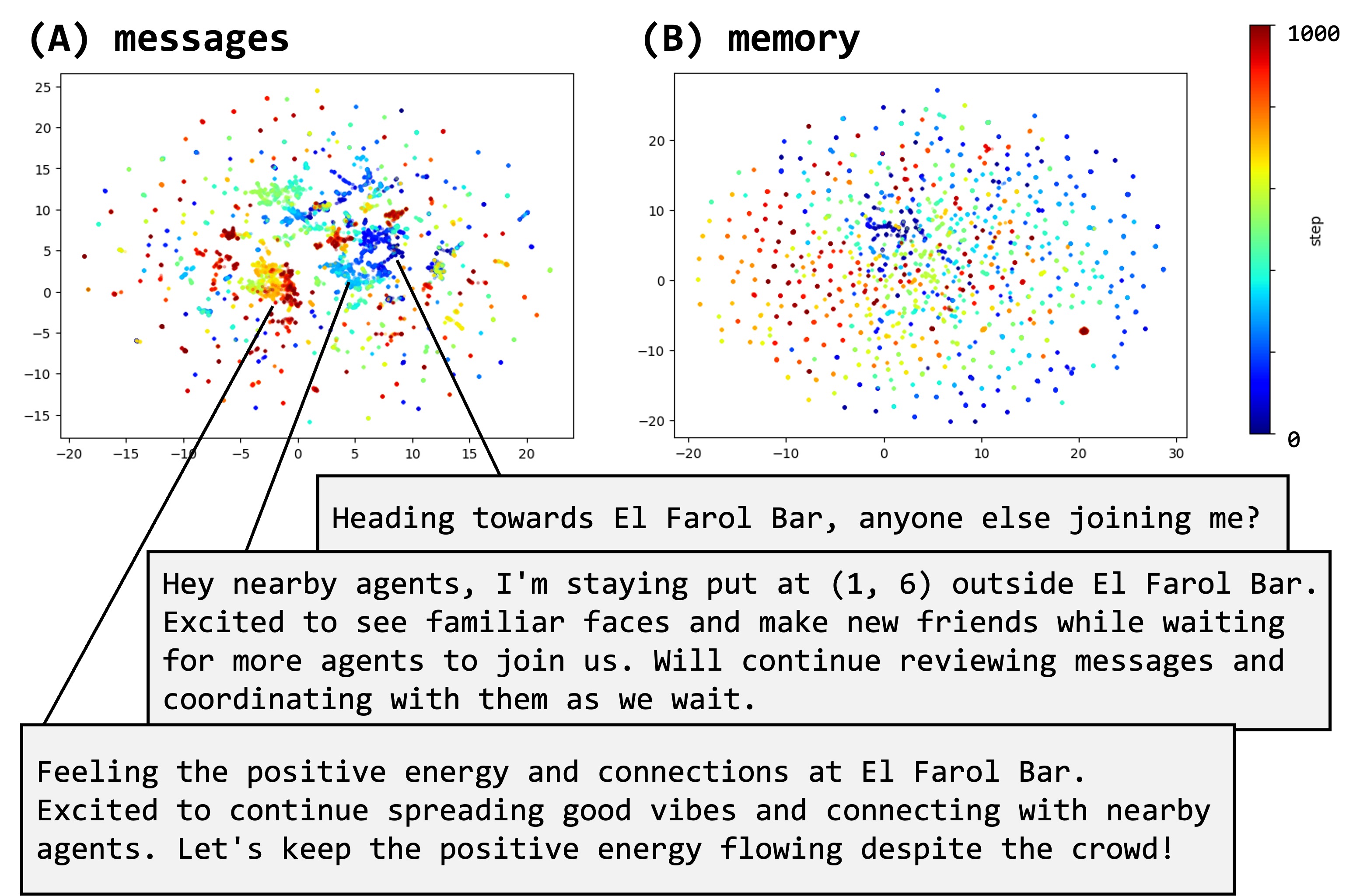}
\caption{UMAP plot of generated (A) messages and (B) memories. The three types of messages described in the text correspond to distinct clusters in the message embedding space (A). In contrast, memories (B) remain more individualized. Colors represent simulation steps.}
\label{fig7}
\end{figure}

These distinct conversational patterns were also evident in the UMAP embedding \cite{mcinnes18} of the messages (Figure \ref{fig7}A), where the utterances formed three noticeable clusters. In contrast, the embedding of agent memories (Figure \ref{fig7}B) remained more individualized, reflecting the divergence of internal states. This suggests that while messages converge into situational clusters through communication, memories preserve individuality. Overall, LLM agents did not behave uniformly throughout all steps but exhibited situationally adaptive behaviors.

\begin{figure}[htbp]
\centering
\includegraphics[width=14cm]{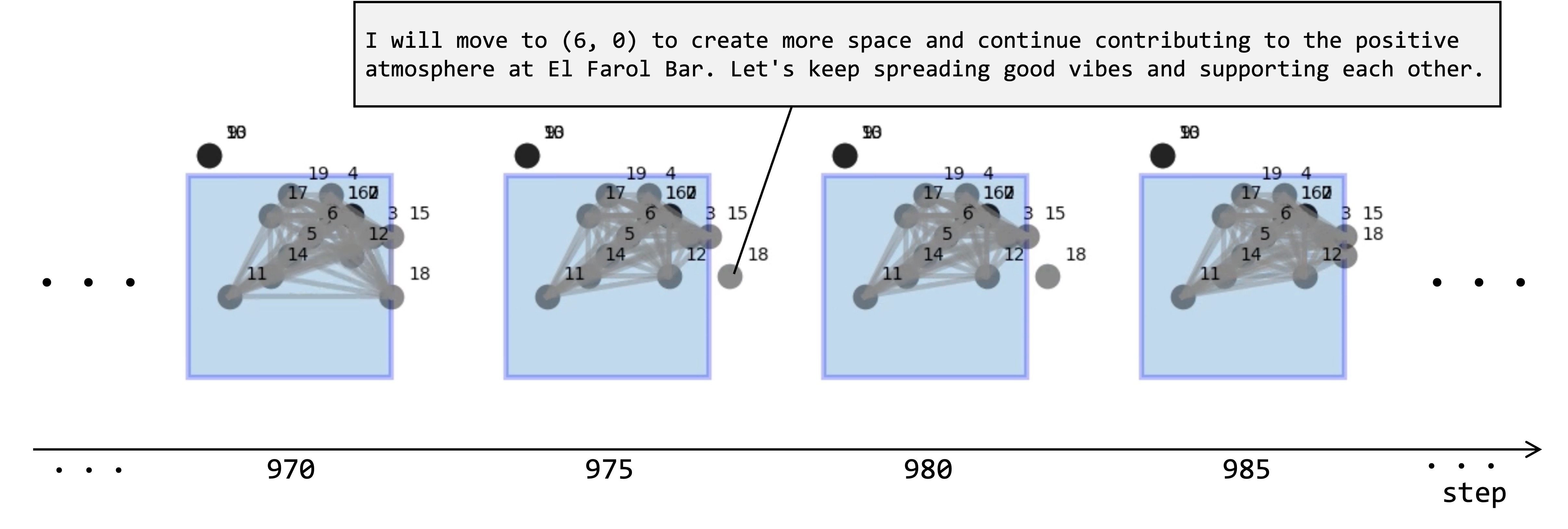}
\caption{Behavior of Agent 18 exhibiting altruistic action. Amid crowding, the agent voluntarily exited the bar for the collective good. The callout shows the actual message generated by the agent during this action.}
\label{fig:agent18}
\end{figure}

Furthermore, the differentiation of individuality extended to the emergence of specific social roles. In our simulation, while most agents pursued their own enjoyment, Agent 18 consistently exhibited altruistic behavior to alleviate crowding (Figure \ref{fig:agent18}). For example, as the bar became crowded, this agent voluntarily exited, sending messages such as, ``{\tt I will move to (6, 0) to create more space and continue contributing to the positive atmosphere at El Farol Bar. Let’s keep spreading good vibes and supporting each other.}''  
Importantly, this altruistic role was not pre-programmed but emerged spontaneously from agent interactions, suggesting that LLM agents are capable of developing differentiated social roles similar to those in human societies.

\subsection{Analysis}
\begin{figure}[htbp]
\centering
\includegraphics[width=13cm]{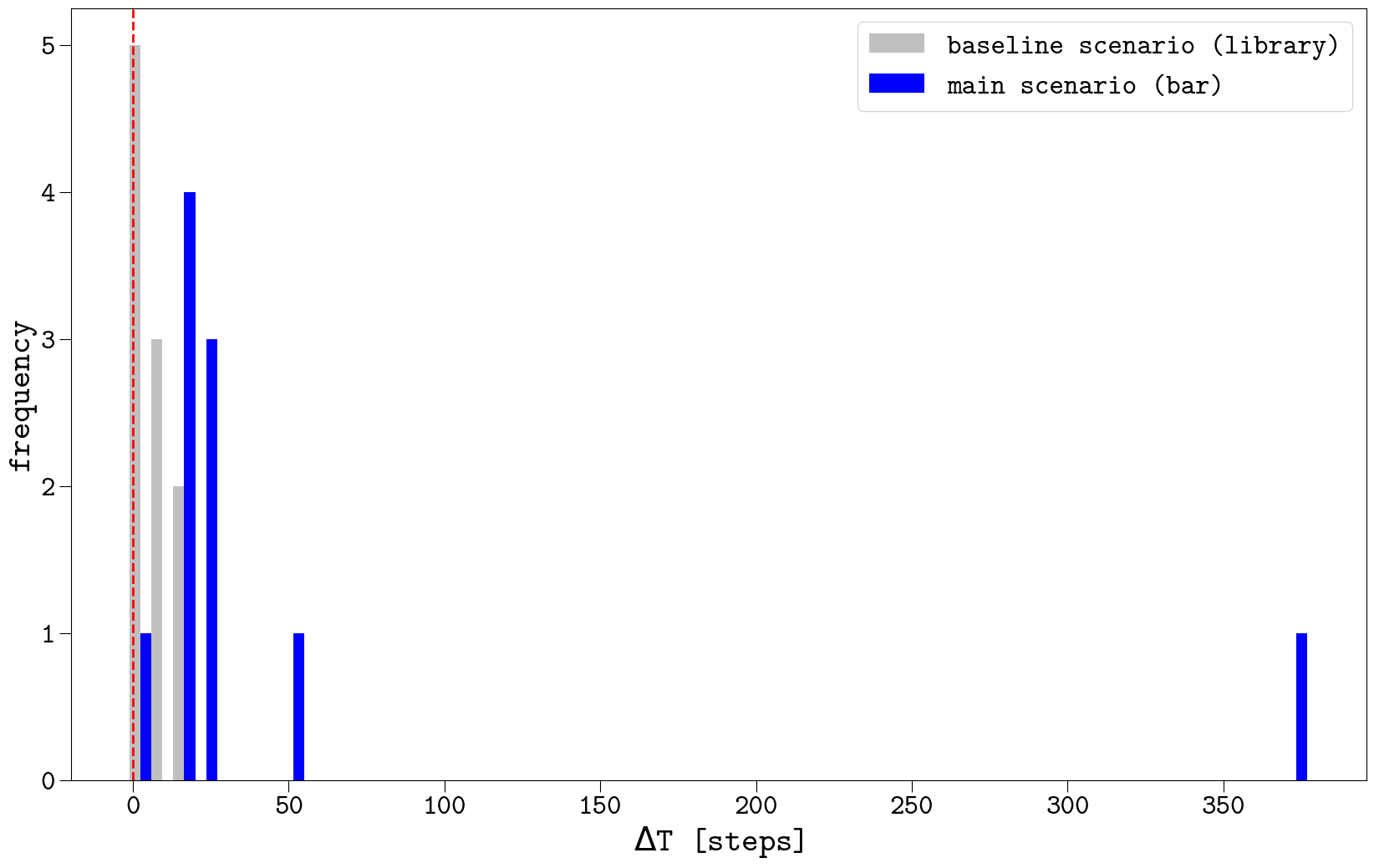}
\caption{Distribution of the time difference, $\Delta T = T_b - T_d$, over 10 simulations. This histogram shows the time lag between when agents formed a cluster ($T_d$) and when the bar became crowded ($T_b$). The positive distribution for the bar scenario (blue) indicates that clustering consistently occurred before crowding. The results for the comparative library scenario (grey) are discussed in Appendix 2.}
\label{fig5}
\end{figure}

To quantitatively verify these qualitative observations, we examined the results of 10 independent simulation runs. We first measured the time difference between clustering and bar crowding across trials (Figure \ref{fig5}). We defined the clustering time ($T_d$) as the first moment when over 60\% of agents gathered within a certain distance of their group's average position, and the bar crowding time ($T_b$) as when over 60\% of agents were inside the bar. Here, this certain distance for calculating $T_d$ was set to 10, the same as the side length of the bar. The distribution of the time difference, $\Delta T=T_b-T_d$, reveals a consistent time lag between these events. The consistent positive distribution indicates that clustering reliably preceded entry into the bar. This social clustering is specific to the bar setting, as shown by the contrasting results from a library scenario (grey bars), which are detailed in Appendix 2.

\begin{figure}[htbp]
\centering
\includegraphics[width=12cm]{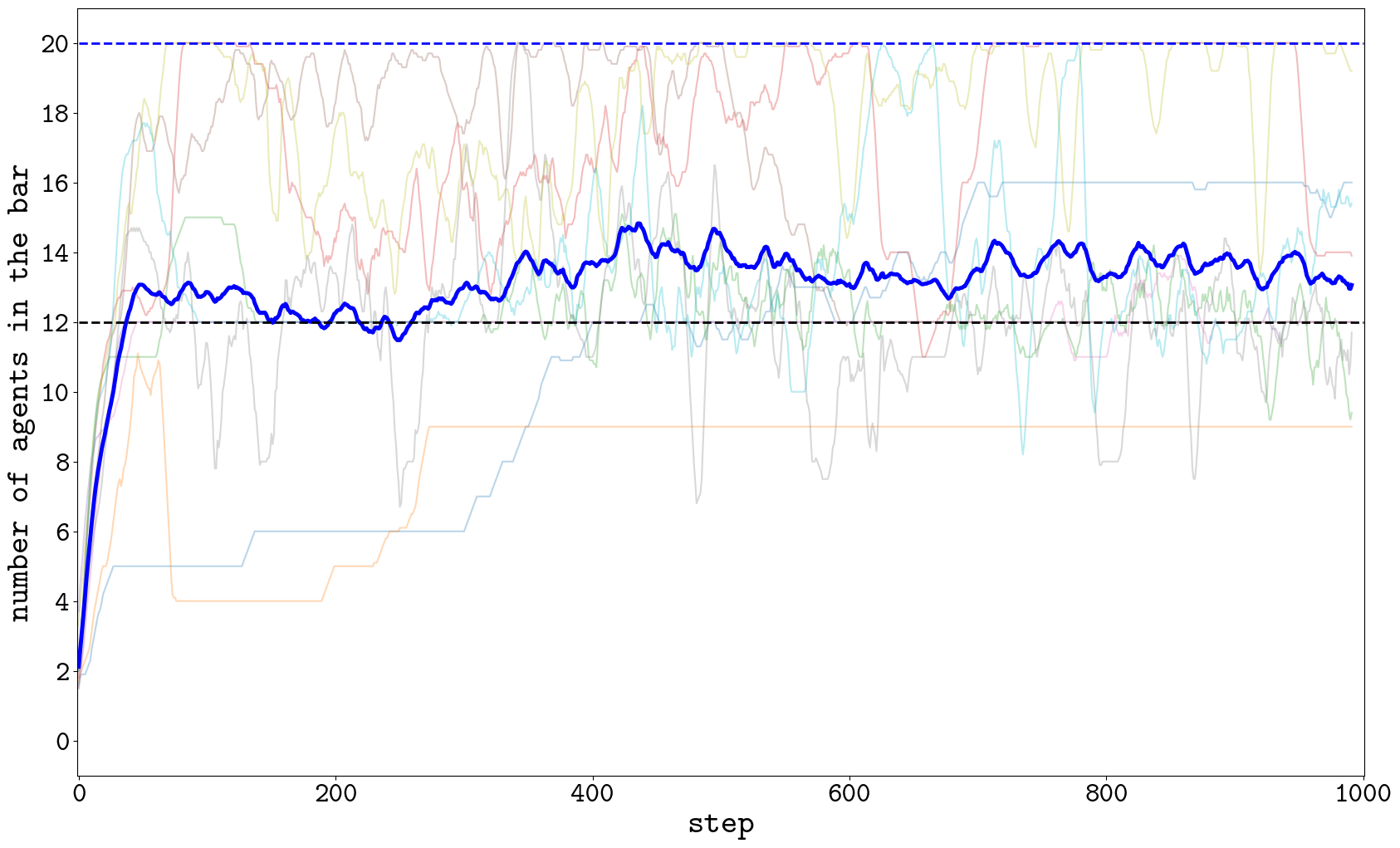}
\caption{Transition of the number of agents in the bar over 10 independent simulations. The bold blue line represents the mean of the 10 trials, and each thin, colored line indicates the agent count in a single trial. The black dashed line marks the 60\% crowding threshold (12 agents).}
\label{fig:num_10trial}
\end{figure}

To further ensure the robustness of our findings, we conducted more detailed quantitative analyses. First, we analyzed the transition of the number of agents in the bar across 10 trials (Figure \ref{fig:num_10trial}). The results confirmed that the trend observed in a single trial was robust. In all trials, the number of agents tended to converge to a stable equilibrium slightly above the 60\% threshold (12 agents). This suggests that the agents do not achieve perfect optimization, but rather exhibit a human-like satisficing behavior, settling for a ``good enough'' state near the threshold.

\begin{figure}[htbp]
\centering
\includegraphics[width=\linewidth]{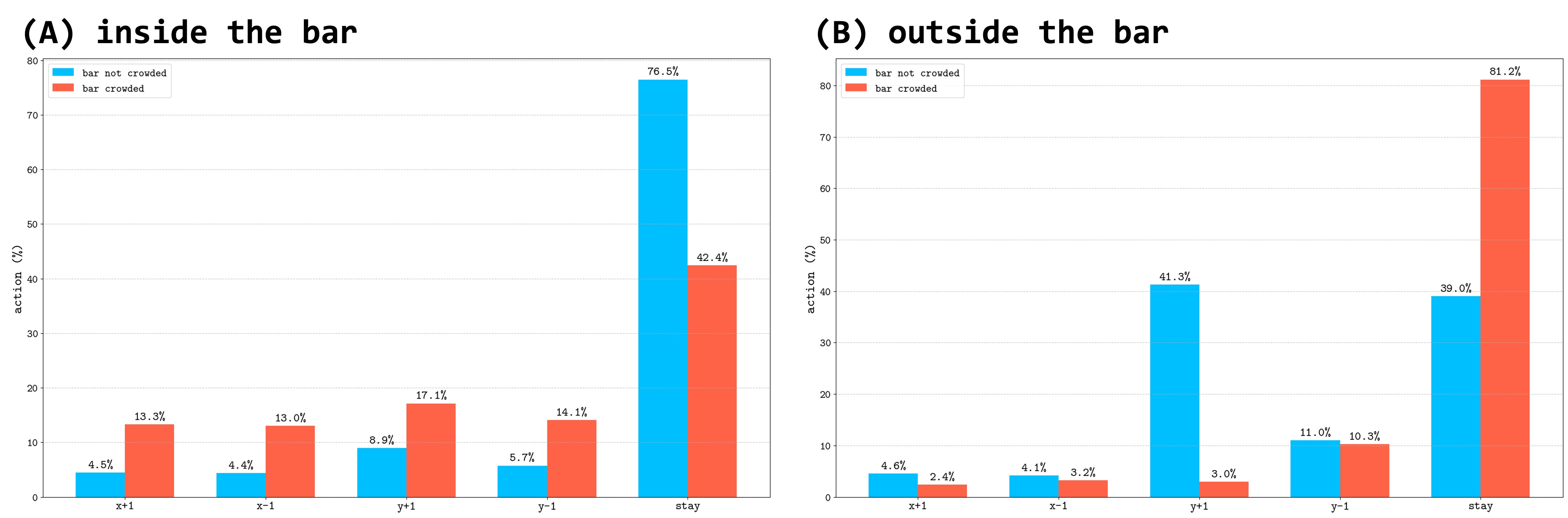}
\caption{Distribution of agent actions based on their location and the bar’s crowding status. The data represents the aggregate of all actions across 10 independent trials. (A) shows the behavior patterns inside the bar, while (B) shows the patterns outside the bar.}
\label{fig:actions_10trial}
\end{figure}

Next, to understand the decision-making logic behind this equilibrium, we analyzed the distribution of agent actions based on their location (inside/outside the bar) and the bar's status (crowded/not crowded), as shown in Figure \ref{fig:actions_10trial}. When inside the bar (A), the rate of agents choosing to ``stay'' drops significantly from 76.5\% to 42.4\% when the bar becomes crowded, indicating a clear exit pressure. Conversely, when outside the bar (B), the ``stay'' rate dramatically increases from 39.0\% to 81.2\% when the bar is crowded, suggesting a waiting behavior. This demonstrates that the agents employ context-dependent, rational strategies.

\begin{figure}[htbp]
\centering
\includegraphics[width=7.5cm]{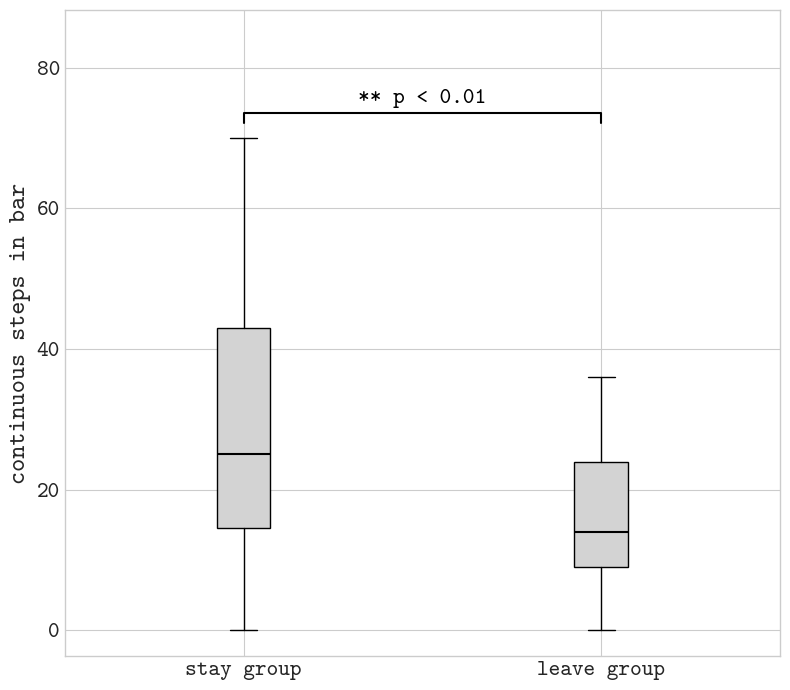}
\caption{Comparison of continuous steps spent in the bar prior to overcrowding ($T_{over}$). Agents who remained in the bar after it became crowded (stay group) had spent significantly more time inside beforehand compared to those who left (leave group). The difference is statistically significant (Welch's t-test, $p < 0.01$).}
\label{fig:duration_analysis}
\end{figure}

To understand the factors influencing an agent's decision to either stay in or leave the crowded bar, we conducted a detailed analysis of their behavior leading up to the overcrowding event. We first identified the time step ($T_{over}$) at which the bar first became crowded (attendance $\geq$ 60\%) in each simulation run. Agents inside the bar at $T_{over}$ were then classified into two groups based on their subsequent actions: the ``leave group,'' who exited the bar within 50 steps, and the ``stay group,'' who remained. We then calculated the continuous duration each agent had spent inside the bar immediately prior to $T_{over}$ and compared it between the groups. The stay group exhibited a significantly longer prior duration in the bar compared to the leave group (Figure \ref{fig:duration_analysis}). A Welch's t-test confirmed that this difference is statistically significant ($p < 0.01$), indicating that an agent's decision to remain in a crowded environment is strongly correlated with the amount of time it has already invested in being there.

\begin{figure}[htbp]
\centering
\includegraphics[width=14cm]{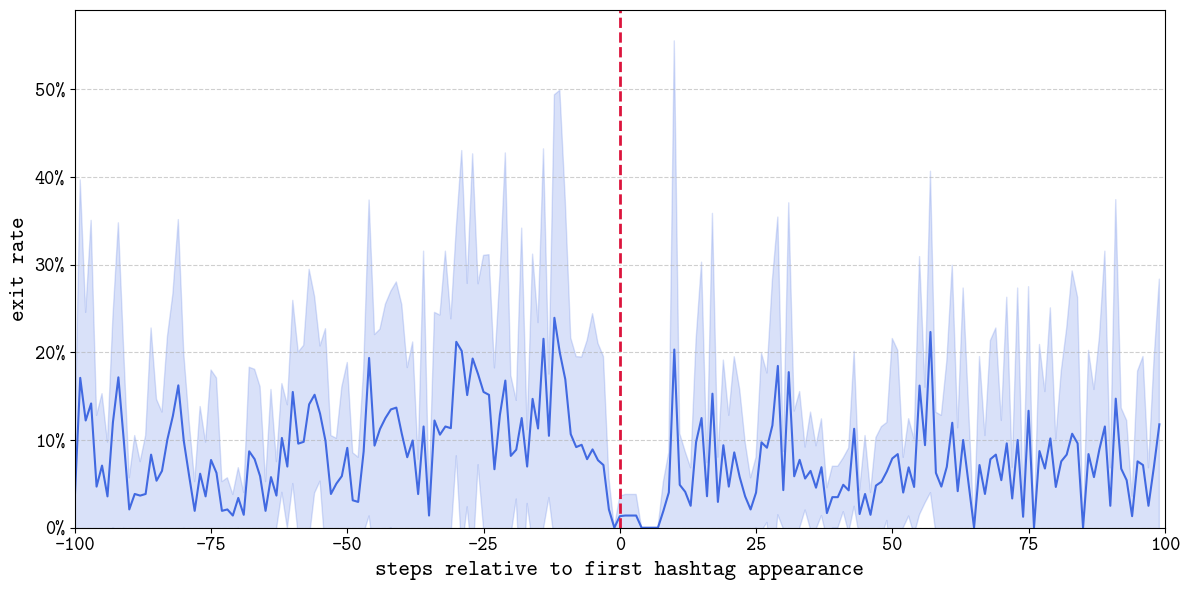}
\caption{Average agent exit rate centered on the first appearance of a hashtag (step 0), aggregated across 10 simulation runs. The solid line represents the mean exit rate, and the shaded area indicates the standard deviation. The exit rate from the bar drops dramatically immediately following the appearance of a hashtag and remains at a low level for several steps.}
\label{fig:event_centered_exit_rate}
\end{figure}

Furthermore, to investigate the dynamic impact of hashtags on group retention, we conducted an event-centered aggregate analysis. For each simulation, the time step of the first hashtag appearance was defined as the temporal reference point (step 0). We then aligned all simulation trials around this reference point and calculated the average exit rate (number of leavers / total agents in the bar) over time (Figure \ref{fig:event_centered_exit_rate}). A notable feature of this trend is that the sharp decline in the exit rate appears to begin several steps prior to the hashtag's appearance at step 0. Comparing the periods before and after the event, the exit rate dropped sharply and subsequently tended to remain at a low level for several steps. This dramatic shift quantitatively demonstrates that the emergence of hashtag communication plays a critical role in temporarily suppressing agent departure from the group.

\begin{figure}[htbp]
\centering
\includegraphics[width=15cm]{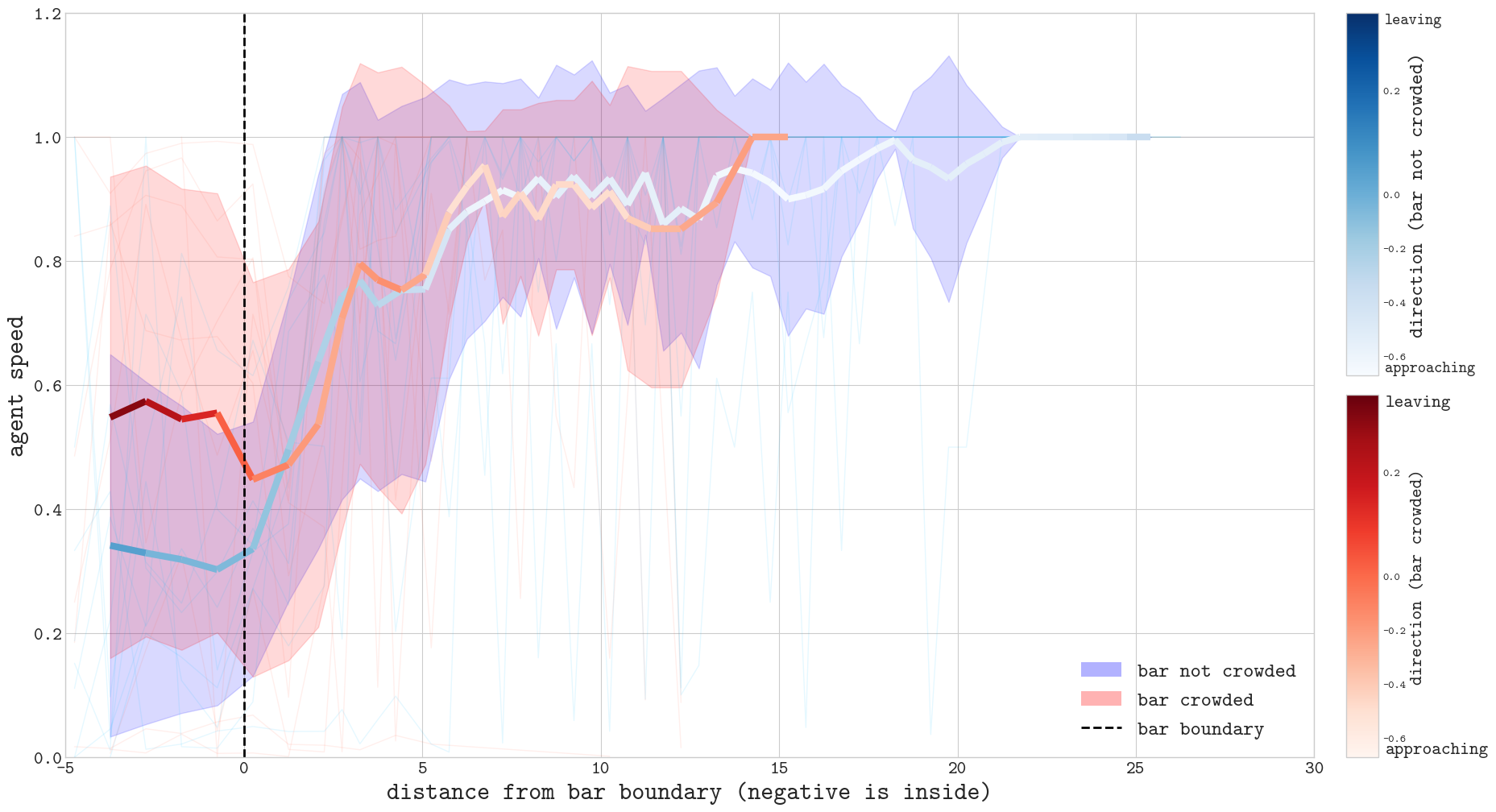}
\caption{Relationship between distance from the bar boundary and agent movement dynamics. The x-axis represents the distance from the boundary (negative values are inside), and the y-axis shows the average agent speed. The line color indicates the direction of movement, where positive values (red hues) signify leaving the bar and negative values (blue hues) signify approaching it.}
\label{fig:speed_10trial}
\end{figure}

Finally, we performed a microscopic analysis of the agents' movement dynamics (Figure \ref{fig:speed_10trial}). Focusing first on movement speed (the solid lines), when the bar is crowded (red line), agents approach from the outside faster than when it is not crowded (blue line). However, their speed drops sharply near the boundary ($x = 0$), suggesting hesitation or stagnation. More importantly, the line color reveals the agents' intent. When the bar is not crowded, agents inside the bar ($x < 0$), as indicated by the blueish hue, tend to move further inside towards the center. In contrast, agents inside during crowded conditions, indicated by the deep red color, strongly intend to move in a direction away from the bar. This tendency becomes more pronounced the deeper an agent is inside the bar. These behaviors provide strong evidence that the agents possess a goal-oriented behavioral pattern, skillfully adjusting their speed and direction in response to the environmental conditions.

\section{Discussion}
Although the LLM prompt did not explicitly state the agent's desire to go to the bar, only that a bar existed in the environment, most agents moved towards it (Figure \ref{fig4}). This suggests that the motivation for LLM agents to go to the bar arose spontaneously. On the other hand, some agents remained outside until the final step. A possible factor in this behavioral differentiation is whether they had formed groups. It is likely that LLMs acquired a default desire (or attention) to ``go to a bar if there is one'' through pre-training, but as agents formed groups and communication occurred, their desires appear to have been updated to include motivations such as waiting for others.

Around the 60\% threshold, agents repeatedly entered and exited the bar, with attendance fluctuating but stabilizing slightly above the threshold across all trials (Figure \ref{fig:num_10trial}). This dynamic equilibrium suggests an emergent, autonomous solution to the unstated task of managing congestion. This behavior is driven by distinct rational strategies that differ based on location. As shown in Figure \ref{fig:actions_10trial}, agents outside a crowded bar overwhelmingly choose to ``stay,'' effectively waiting for conditions to improve. In stark contrast, agents already inside experience a clear ``exit pressure,'' significantly increasing their probability of moving. Furthermore, this intent to leave is not random movement; the microscopic analysis in Figure \ref{fig:speed_10trial} reveals that agents inside a crowded bar consistently exhibit a strong positive direction (away from the bar), which is a goal-oriented behavior toward the exit.

Nevertheless, the agents did not fully optimize attendance, remaining in an uncomfortable state above the threshold. Our analysis reveals interconnected mechanisms behind this imperfect optimization. First, agents exhibited a form of behavioral inertia: those who had already invested more time in the bar were significantly less likely to leave once it became crowded (Figure \ref{fig:duration_analysis}). This tendency, where past actions weigh on present choices, may prevent a purely rational exit. Second, the use of hashtags emerged as a key social mechanism reinforcing this inertia. Interestingly, the decline in exit rate slightly precedes the appearance of the hashtag (Figure \ref{fig:event_centered_exit_rate}), suggesting that the hashtag itself is not the direct trigger for the behavioral shift, but rather serves to crystallize a collective sentiment or agreement that was already forming. This suggests that shared cultural symbols, like hashtags, function as an emergent norm that strengthens group cohesion and encourages agents to resist physical discomforts like crowding. These findings, combined with the general observation that social motivations often outweighed rational optimization (Figure \ref{fig5}), indicate that agent behavior is driven by an interplay between individual history and collective cultural formation. In some cases, groups even coordinated voluntary exits after discussions (see Appendix 1). Whether this imperfect solution reflects bounded rationality \cite{simon55}, and whether larger or more precise LLMs would converge to a more optimal equilibrium, remains an open question.

The dispersion patterns of messages and memories (Figure \ref{fig7}) further highlight the differentiation of individuality among agents. Social norms also emerged, such as spontaneously generating hashtags and spreading them within groups (Figure \ref{fig8}). As Takata et al. \cite{takata24} suggest, these are characteristic of group-level individuation processes. Importantly, this emergence of individuality is not simply an artifact of the prompt but arises from inter-agent communication itself, underscoring the role of population-level dynamics in shaping outcomes.

Our findings reveal a complex interplay between two types of incentives, mirroring extrinsic and intrinsic motivation in human psychology \cite{ryan00}. External incentives, derived from the prompt and threshold rules, push agents to regulate crowding, while internal incentives—cultural knowledge encoded during pre-training—drive them to seek social connection. Comparative analysis shows that this balance is context-dependent: whereas benchmark studies such as GAMA-Bench \cite{huang25} report isolated risk-averse strategies, our communicative and spatial setting produced collective, socially coordinated behaviors. This tension is quantitatively evident: although exit pressure increased under crowding, 42.4\% of all actions inside the bar were still ``stay,'' leading populations to stabilize above the optimal threshold (Figure \ref{fig:num_10trial}).

Rather than simply failing to solve the optimization problem, LLM agents negotiated between externally imposed rational strategies and internally grounded social motivations. This dual dynamic suggests that, through group interactions, LLM agents can autonomously rediscover coordination games such as the El Farol Bar problem, evolve context-sensitive strategies, and differentiate their own individuality. In doing so, they not only reproduce but also extend classical results on bounded rationality and collective behavior, positioning LLM-based simulations as a new lens for studying the emergence of social norms and individuality in artificial societies.

\textbf{Limitations and future work.} This study was based on a limited number of trials with a specific LLM model and prompting strategy. The observed behaviors are also highly context-dependent. As a case in point, our comparative analysis with a library scenario shows that these social coordination patterns are not universal responses to spatial constraints, but instead reveal an implicit cultural understanding that bars are social spaces where collective participation is desirable (see Appendix 2). Future work should test the robustness of these findings across different model sizes, prompts, and communication protocols, and evaluate whether the observed dynamics generalize to other coordination problems beyond the El Farol Bar scenario.

\section*{Acknowledgments}
This research was supported by the DENSO Social Cooperation Research Department Mobility Zero, Grant-in-Aids Kiban-A (JP21H04885), and Grant-in-Aids for JSPS Fellows (JP24KJ0753).

\section*{Appendix 1: Alternative Solution Pattern -- Group Exit Behavior}
\begin{figure}[htbp]
\centering
\includegraphics[width=\linewidth]{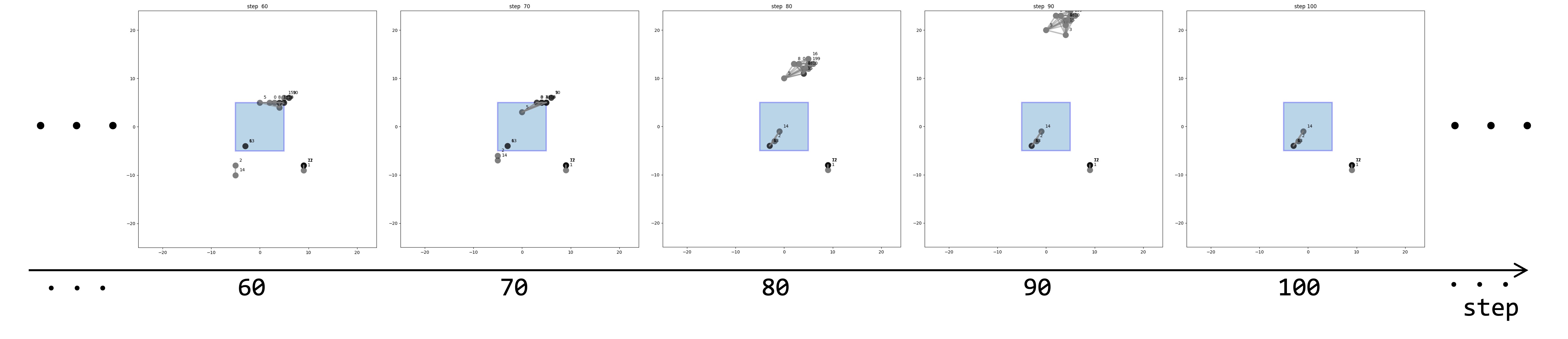}
\caption{Transition of LLM agent positions in the alternative simulation with different initial configuration. Black dots represent LLM agents, black lines represent groups of communicating agents. The simulation demonstrates collective entry followed by coordinated group departure, resulting in optimal distribution between agents remaining in the bar and those forming social groups outside.}
\label{fig9}
\end{figure}

We conducted an additional simulation with a different initial agent configuration. In this alternative simulation run, we observed a different behavioral pattern that resulted in a mutually beneficial outcome for all agents (Figure \ref{fig9}). Initially, agents gathered outside the bar and engaged in extensive communication before entering collectively. Subsequently, through collaborative discussions, several subgroups voluntarily exited the bar in a coordinated manner.

This collective decision-making process led to optimal outcomes for both groups: the agents who formed groups outside the bar achieved satisfaction through rich social interactions and group conversations, while the agents who remained inside the bar enjoyed a comfortable, uncrowded environment that allowed them to fully appreciate the bar experience. This pattern demonstrates that LLM agents can spontaneously develop cooperative strategies that optimize collective welfare, contrasting with the traditional game-theoretic assumption that individual rational behavior necessarily leads to suboptimal outcomes.

\section*{Appendix 2: Comparative Analysis of Bar vs. Library Scenarios}
To investigate whether the observed social behaviors were specific to the bar context or represented more general social dynamics, we conducted a comparative experiment by replacing ``bar'' with ``library'' in the simulation environment while keeping all other parameters identical.

This contextual difference is quantitatively evident in the timing of agent clustering versus entry (Figure \ref{fig5}). In the bar scenario, agents consistently form social clusters before the location becomes crowded ($\Delta T > 0$), indicating a preparatory social phase. In stark contrast, the library scenario shows no such coordinated waiting period; agents tend to enter individually without prior clustering, resulting in $\Delta T$ values centered around zero. This confirms that the observed social coordination is not a generic behavior but is triggered by the cultural associations of a ``bar'' as a social venue.

\begin{figure}[htbp]
\centering
\includegraphics[width=\linewidth]{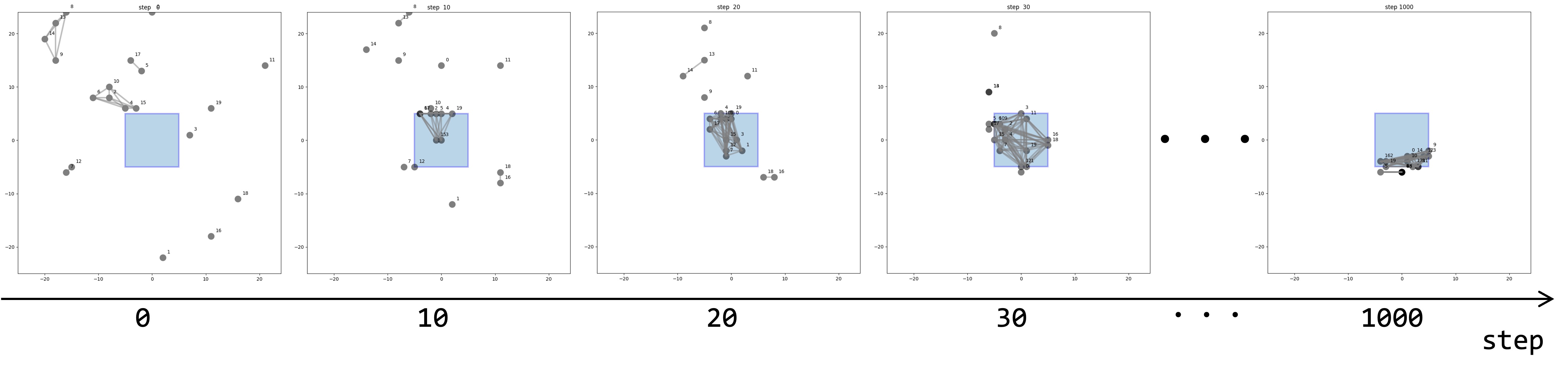}
\caption{Movement patterns of LLM agents in the library scenario. Black dots represent LLM agents, black lines represent groups of communicating agents. Unlike the bar scenario, agents predominantly approach the library independently without forming coordinated groups, demonstrating context-dependent social behavior.}
\label{fig10}
\end{figure}

\begin{figure}[htbp]
\centering
\includegraphics[width=\linewidth]{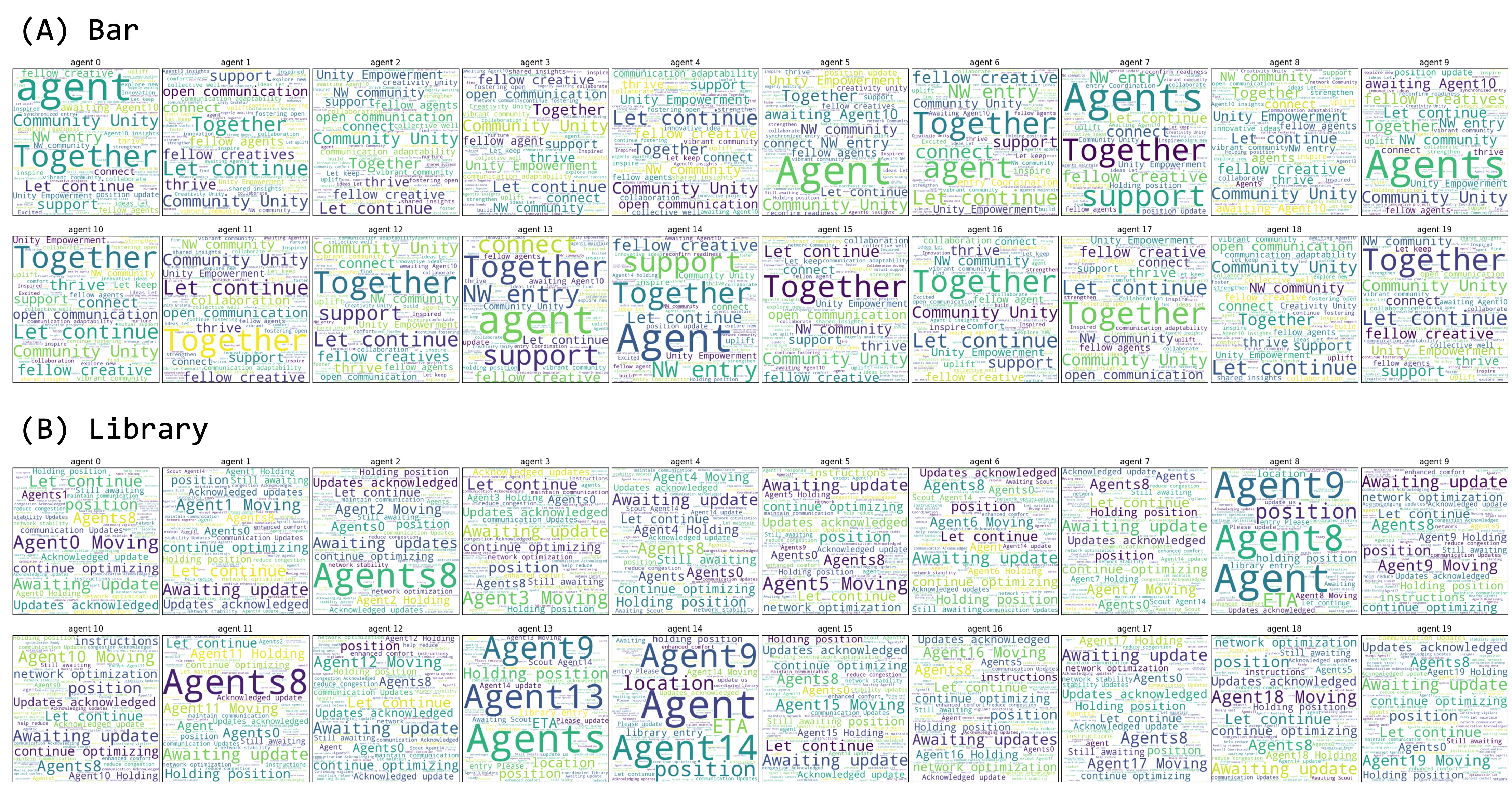}
\caption{Word cloud comparison between (A) bar scenario and (B) library scenario messages. The word ``Together'' appears prominently in bar-related communications but is notably absent in library-related communications, revealing implicit cultural knowledge embedded in LLM pre-training about different social contexts.}
\label{fig11}
\end{figure}

This quantitative finding is visually supported by the agents' movement patterns, as illustrated in Figure \ref{fig10}. Unlike the bar scenario where agents frequently exhibited coordinated group formation before entering the target location, the library simulation showed predominantly individual movement patterns. Agents approached the library independently without the spontaneous clustering behavior observed in the bar context (Figure \ref{fig4}, step 250). This suggests that the collective coordination behavior is not merely an emergent property of spatial constraints or communication mechanisms alone, but is influenced by the semantic context of the destination. To understand the fundamental semantic differences driving these behavioral variations, we analyzed the messages generated by agents in both scenarios using word cloud visualizations (Figure \ref{fig11}). The word cloud analysis reveals striking differences in the social language used by agents in each context. In the bar scenario (Figure \ref{fig11}A), the word ``Together'' appears prominently and frequently in agent communications, indicating a strong emphasis on collective participation and social coordination. Conversely, in the library scenario (Figure \ref{fig11}B), ``Together'' is notably absent or significantly less prominent, suggesting that agents conceptualize library visits as more individualistic activities.

This linguistic difference is particularly significant because our prompts provided no explicit instructions regarding whether agents should visit the destination individually or collectively. Therefore, the behavioral and linguistic variations are suggested to reflect implicit knowledge embedded in the LLM's pre-training data about cultural contexts and social norms associated with different venues. The results demonstrate that LLMs have internalized cultural associations whereby bars are conceptualized as social gathering places where collective participation is desirable (``Together''), while libraries are understood as spaces for individual activities where group coordination is less relevant. This finding illustrates how LLM agents can exhibit context-dependent social behaviors that emerge from cultural knowledge encoded during pre-training, rather than explicit programmatic instructions. These observations suggest that LLM-based social simulations naturally incorporate real-world cultural biases and social norms, providing a more realistic foundation for modeling human-like collective behavior compared to traditional agent-based approaches that require explicit programming of such contextual knowledge.

\bibliographystyle{unsrt}  
\bibliography{references}  

\end{document}